\documentclass[chicago,apj]{emulateapj-rtx4}
\slugcomment{{\sc Published in ApJ:} 2010, 722, 20}
\usepackage{lscape}
\newcommand{\msun}{\mbox{$M_\odot$}}

\newcommand{\hst}{\emph{HST}}
\newcommand{\chandra}{\emph{Chandra}}
\newcommand{\rosat}{\emph{\small ROSAT}}

\newcommand{\br}{\mbox{$B\!-\!R$}}
\newcommand{\ha}{\mbox{H$\alpha$}}
\newcommand{\hr}{\mbox{$\ha\!-\!R$}}

\shortauthors{Cohn et al.}
\shorttitle{Faint \chandra\ Sources in NGC~6397}

\begin{document}


\title{Identification of Faint \chandra\ X-ray Sources in the
  Core-Collapsed Globular Cluster NGC~6397: Evidence for a Bimodal
  Cataclysmic Variable Population}

\author{Haldan N. Cohn \& Phyllis M. Lugger} 
\affil{Department of Astronomy, Indiana University, 727 E. Third St.,
Bloomington, IN 47405; cohn@astro.indiana.edu, lugger@astro.indiana.edu}

\author{Sean M. Couch}
\affil{Department of Astronomy, University of Texas, RLM 15.202A,
Austin TX 78712; smc@astro.as.utexas.edu}

\author{Jay Anderson}
\affil{Space Telescope Science Institute, Baltimore, MD 21218;
jayander@stsci.edu}

\author{Adrienne M. Cool}
\affil{Department of Physics and Astronomy, San Francisco State
University, 1600 Holloway Avenue, San Francisco, CA 94132;
cool@sfsu.edu}

\author{Maureen van den Berg}
\affil{Harvard-Smithsonian Center for Astrophysics, 60 Garden St., Cambridge, MA 02138;
maureen@head.cfa.harvard.edu} 

\author{Slavko Bogdanov}
\affil{Department of Physics, 
McGill University,
Ernest Rutherford Physics Building 226,
Montreal, PQ H3G 1A9
Canada;
bogdanov@physics.mcgill.ca}

\author{Craig O. Heinke}
\affil{Department of Physics, University of Alberta, Edmonton, AB T6G
2G7, Canada; cheinke@phys.ualberta.ca }

\and

\author{Jonathan E. Grindlay} 
\affil{Harvard-Smithsonian Center for Astrophysics, 60 Garden St., Cambridge, MA
02138; josh@cfa.harvard.edu }

\begin{abstract}
We have searched for optical identifications for 79 \chandra\ X-ray
sources that lie within the half-mass radius of the nearby,
core-collapsed globular cluster NGC 6397, using deep Hubble Space
Telescope Advanced Camera for Surveys Wide Field Channel imaging in
\ha, $R$, and $B$\@.  Photometry of these images allows us to classify
candidate counterparts based on color-magnitude diagram location.  In
addition to recovering nine previously detected cataclysmic variables
(CVs), we have identified six additional faint CV candidates, a total
of 42 active binaries (ABs), two millisecond pulsars (MSPs), one
candidate active galactic nucleus, and one candidate interacting
galaxy pair. Of the 79 sources, 69 have a plausible optical
counterpart.

The 15 likely and possible CVs in NGC 6397 mostly fall into two
groups: a brighter group of six for which the optical emission is
dominated by contributions from the secondary and accretion disk, and a
fainter group of seven for which the white dwarf dominates the optical
emission. There are two possible transitional objects that lie between
these groups.  The faintest CVs likely lie near the minimum of the CV
period distribution, where an accumulation is expected.  The spatial
distribution of the brighter CVs is much more centrally concentrated
than those of the fainter CVs and the active binaries.  This may
represent the result of an evolutionary process in which CVs are
produced by dynamical interactions, such as exchange reactions, near
the cluster center and are scattered to larger orbital radii, over
their lifetimes, as they age and become fainter.

\end{abstract}  


\section{Introduction}

Gravitational interactions involving ``hard'' binary stars (i.e.\
those with orbital velocities that exceed the local velocity
dispersion) provide the energy source that drives globular cluster
evolution from the time that the cluster core approaches collapse
\citep{Hut92,Heggie03}.  These interactions have a profound effect on
the long-term fates of both the cluster and the binaries.  Among the
hardest binaries in clusters are compact, accretional X-ray sources.
Bright low-mass X-ray binaries (LMXBs), which contain a neutron star
primary, are strongly overabundant in dense clusters and thus almost
certainly have a dynamical origin there
\citep[see][]{Pooley03,Ivanova08}.  There is also growing evidence
that a majority of the more numerous cataclysmic variables (CVs) in
clusters may also have a dynamical origin, rather than representing a
primordial population \citep{Pooley06}.  This is a complex issue,
given the many possible channels for the formation of CVs in clusters,
which include: the evolution of primordial binaries, mediated by
hardening interactions; exchange interactions between singles and
primordial binaries, possibly involving multiple exchanges; tidal
captures which do not result in mergers; and three-body interactions
\citep[see][]{Ivanova06}.  Deep surveys of compact binary populations
in globular clusters will help to clarify the relative importance of
these CV formation mechanisms.

NGC 6397 is the second closest globular cluster \citep[$d = 2.3~{\rm
    kpc}$;][]{Strickler09} and is by far the nearest cluster to have
undergone core collapse.  It thus provides a special laboratory for
studying the formation and evolution of compact binary systems in the
extreme environment of a collapsed core.  NGC~6397 has been a key
target for \rosat, Hubble Space Telescope (\hst), and Chandra X-ray
Observatory (\chandra) studies of compact binaries in clusters.
\citet{Cool95} first detected cataclysmic variables in NGC~6397, as
counterparts to faint \rosat\ sources, using
\hst\ Wide-Field/Planetary Camera~1 \ha\ and $R$-band imaging to
select \ha-excess objects.  \citet{Grindlay95} confirmed the power of
the \hr\ photometric technique, providing \hst\ Faint Object
Spectrograph verification of the first three CV identifications in
NGC~6397.  \citet{Grindlay01} carried out the first
\chandra\ observations of NGC~6397, which detected a rich population
of 25 sources within the cluster half-mass radius \citep[$r_h$ =
  2.33\arcmin;][]{Harris96}.  WFPC2 imaging in \ha, $R$, $V$, and $I$
has led to the identification of many of these sources, including nine
CVs and a number of active binaries \citep{Taylor02,Grindlay06}.
Since the study by \citet{Taylor02} was carried out, we have obtained
considerably deeper \chandra\ and \hst\ imaging of NGC~6397.  The
results of the \chandra\ imaging, which detected 79 sources within the
half-mass radius, are reported by \citet{Bogdanov10}.  These X-ray sources,
designated by a prefix U, are listed in Table~\ref{t:counterparts}.
In this paper, we report a search for optical counterparts to these
sources using a deep \hst\ ACS/WFC\footnote{Advanced Camera for
  Surveys, Wide Field Channel} dataset.  We describe the data, the
analysis method, and the results in the following sections.

\section{Data \label{data}}

The optical imaging used in this study is the \hst\ GO-10257 dataset
(PI: Anderson), which provides deep, highly dithered ACS/WFC imaging
of the central region of NGC~6397 in F435W ($B$), F625W ($R$), and
F658N (\ha).  One of the main motivations of this study was to search
for wobbles in the positions of stars, indicative of massive unseen
companions, such as white dwarfs, neutron stars, or black holes.  To
provide good time sampling, the center of NGC 6397 was imaged over 10
single-orbit epochs, spaced at approximately one-month intervals
between 2004 July and 2005 June.  A by product of this observing
strategy is that we imaged the cluster at many different orientations,
allowing for an exquisite mitigation of detector-dependent photometric
and PSF-related errors.  In addition, the stacked images show round
PSFs with no sign of diffraction spikes.  The dataset consists of 5
short $B$ (13\,s), 5 long $B$ (340\,s), 5 short $R$ (10\,s), 5 long R
(340\,s), and 40 \ha\ (390\,s) exposures.  The short exposures were
designed to provide accurate photometry for stars above the
main-sequence turnoff (MSTO)\@.  Saturation only becomes an issue near
the tip of the giant branch.  With the large number of \ha\ frames,
the PSF sampling is particularly good for this band.

\section{Analysis Method \label{analysis}}

\subsection{Photometry \label{photometry}}

The \hst\ data were analyzed using software based on the program
developed for the ACS Globular Cluster Treasury project, described in
\citet{Anderson08}.  The routine first finds the astrometric and
photometric mapping from each exposure into a master reference frame.
It then searches through the master reference frame one small patch at
a time, identifying stars where a significant number of exposures
indicate coincident detections.  These stars are then measured two
ways: (1) they are measured in each individual exposure where they
could be found, and (2) they are also measured simultaneously in all
the exposures (which provides a better measurement for the fainter
stars).  Stars were measured with spatially variable, library PSFs
(see Anderson \& King ACS ISR
2006-01\footnote{http://www.stsci.edu/hst/acs/documents/isrs/isr0601.pdf})
constructed from the GO-9444 data set (PI-Cool) of $\omega$\,Cen,
using an aperture of whole pixels that was optimized for the
brightness of the star and its particular surroundings.

Since our best coverage was in \ha, we did our star-finding on those
images, and then used the positions found to identify and measure the
stars in the B and R images.  The final photometry we report here
comes from averaging the photometry from the individual stars, with
sigma clipping applied to remove outlying values due to cosmic rays,
defective pixels, etc.  A total of 25004 stars were detected.  For the
$B$ and $R$ bands, the photometry was performed independently for the
short and long frames.  Photometric calibration to the {\small
VEGAMAG} system was adopted from \citet{Strickler09}.  

\begin{figure}
\epsscale{1.2}
\includegraphics[width=3.4in]{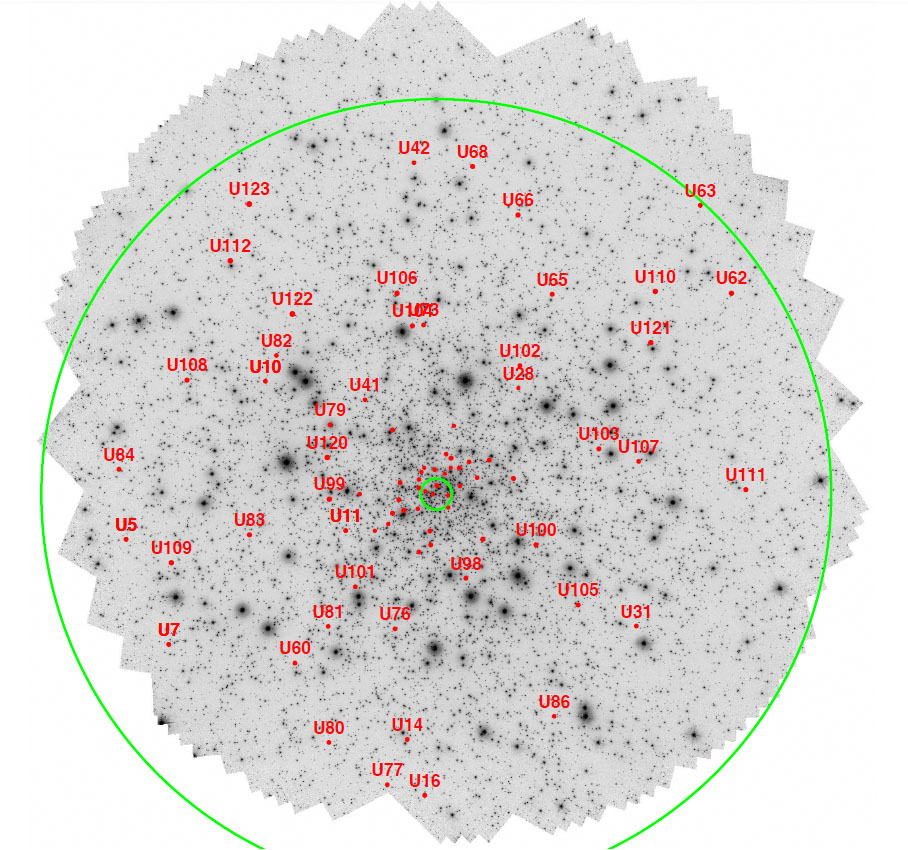}
\figcaption{Stacked \hst\ ACS/WFC in the \ha\ filter of NGC~6397
with \chandra\ source error circles.  North is up and east is to the
left.  The source labels have been omitted for sources within
30\arcsec\ of the cluster center for clarity; these are shown in
Fig.~\ref{f:mosaic_zoom}.  The inner green circle represents the core
radius of 5.5\arcsec\ and the outer green circle represents the
half-mass radius of 2.33\arcmin.  There are a total of 79 sources
detected within the half-mass radius.  
\label{f:mosaic}}
\end{figure}

\begin{figure}
\epsscale{1.2}
\includegraphics[width=3.4in]{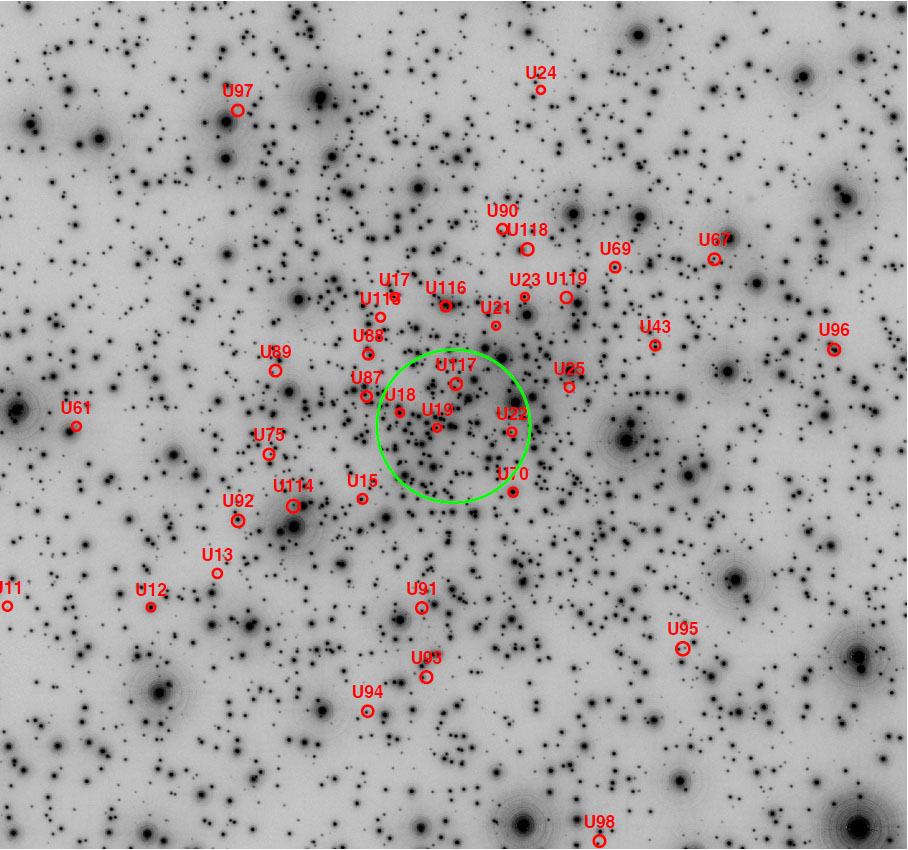}
\figcaption{The inner $1'\times1'$ region of the stacked \ha\ image
  with Chandra source error circles.  The error circles are shown at
  their actual sizes.  The green circle represents the core radius of
  5.5\arcsec.  
\label{f:mosaic_zoom}}
\end{figure}

In addition to the star finding and measuring, we also produced
stacked images of the scene, in a procedure that is akin to the
{\small IRAF/STSDAS} \emph{drizzle} algorithm with {\tt pixfrac} set
to zero.  The stacked images were oversampled by a factor of two, in
order to take advantage of the heavy dithering to increase the
effective resolution.  The resulting supersampled stacked images have
a 12,000$\times$12,000 format and cover an approximately circular
field of diameter 5\arcmin\ with a pixel scale of 0.025\arcsec.
Figures \ref{f:mosaic} and \ref{f:mosaic_zoom} show a combination of
the 40 \ha\ images, together with the error circles for the 79
\chandra\ sources.  The error circles are plotted at their actual
sizes in both figures.  As can be seen in these figures, the
combination of images at various orientations results in a very smooth
symmetric PSF, with a distinct lack of diffraction spikes.

\begin{figure*}
\epsscale{0.8}
\hspace*{0.75in}\includegraphics[width=5.5in]{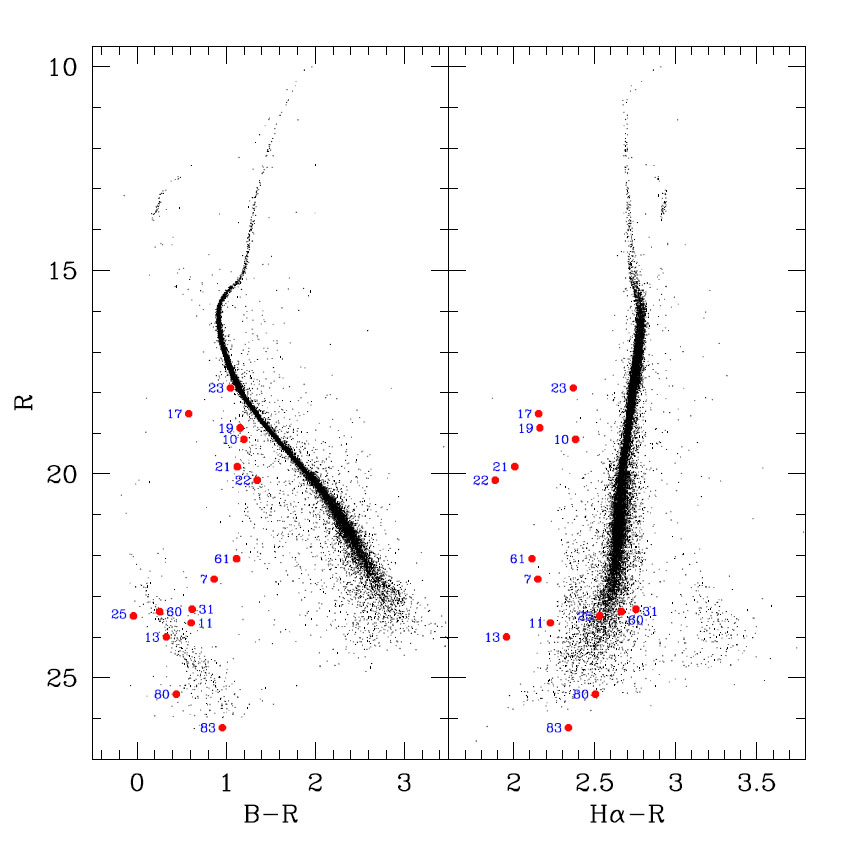}
\figcaption{Color-magnitude diagrams for stars within the half-mass
  radius of NGC~6397 and CV identifications.  The new candidates have
  been selected based on their blue color and \ha\ excess.  Note that
  in the right panel, the bright CVs lie to the \ha-excess side of the
  MS, while the faint CVs lie to the \ha-excess side of the WD clump,
  which itself lies to the \ha-deficit side of the MS.  
\label{f:CMD_CV}}
\end{figure*}

\begin{figure*}
\epsscale{0.8}
\hspace*{0.75in}
\includegraphics[width=5.5in]{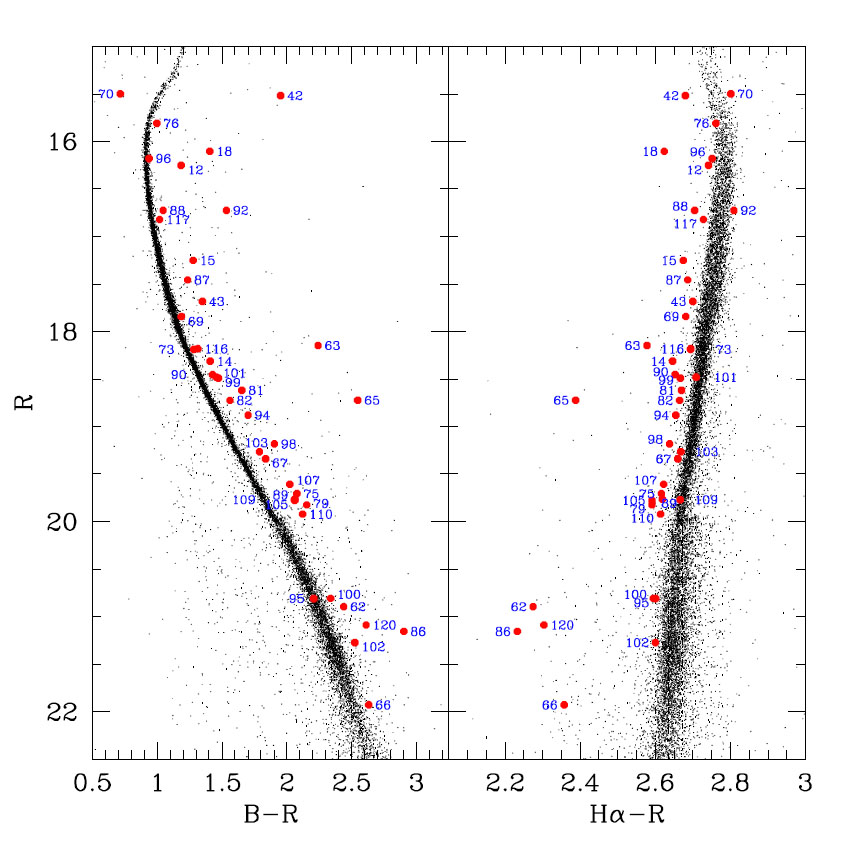}
\figcaption{Color-magnitude diagrams for stars within the half-mass
  radius of NGC~6397 and AB/MSP identifications.  The new
  candidates have been selected based on their red color and generally
  small \ha\ excess.  The object U12 is a known MSP and U18 is a
  likely MSP\@.  The object U70 is a blue straggler.  Eight atypical
  ABs---U42, U62, U63, U65, U66, U86, U92, and U120---are considerably
  redder than typical and/or have larger than typical \ha\ excesses.
\label{f:CMD_AB}}
\end{figure*}

Color-magnitude diagrams (CMDs) were constructed from the photometry
using the $R$ magnitudes, and the \br\ and \hr\ color indices.  The
short-exposure $B$ and $R$ magnitudes were used for stars brighter
than about 2.5 mag below the MSTO, since the long-exposure frames
saturate above this point.  Figures \ref{f:CMD_CV} and \ref{f:CMD_AB}
show the resulting CMDs.  The efficacy of the photometric procedure is
indicated by the tightness of the fiducial sequences.  No
proper-motion cleaning has been performed at this point in the
analysis.  Proper-motion determination is described in
\S\ref{astrometry}.  In \S\ref{source_ID} we will use proper motions
in our source identification.  Unfortunately the quality of the
cluster-field separation depends too sensitively on brightness to
provide a definitive membership determination for all stars.  The
(\br, $R$) CMD reaches deepest for the bluest stars, since the
faintest red main-sequence (MS) stars are below the detection limit in
$B$.  The upper part of the white dwarf (WD) cooling sequence is
clearly detected in the (\br, $R$) CMD, extending to nearly 10 mag
below the MSTO in $R$.  There is also an indication of a second WD
sequence above the primary sequence, which \citet{Strickler09}
interpret as a He WD sequence.  Note that in the (\hr, $R$) CMD, the
WDs lie to the \ha-deficit side of the MS, reflecting the strong
\ha\ absorption lines in WD spectra relative to those of faint MS
stars.  The blue stragglers and the blue horizontal branch stars
similarly lie to the \ha-deficit sides of the subgiant branch and the
giant branch, respectively.

\subsection{Astrometry\label{astrometry}}

The stacked \hst\ ACS/WFC images for each filter were rectified to a
common astrometric system.  We calculated an absolute astrometric
zeropoint for the stacked images relative to the ICRS using a two-step
process.  We first computed a plate solution for an ESO WFI frame
(WFI.2002-06-21T01:45:08.153) using 357 primary astrometric standards
from the USNO UCAC2 catalog.  We then selected 350 secondary
astrometric standards from the WFI frame and used these to compute a
plate solution for our stacked ACS/WFC images.  We determined a
boresight correction for the \chandra\ source coordinates from
\citet{Bogdanov10} by computing the mean offsets between the \hst\ and
\chandra\ coordinates for the six brightest CVs and the millisecond
pulsar (MSP) PSR~J1740-5340.  This resulted in a shift of the
\chandra\ coordinates to the ICRS of $-0.19''\pm0.02''$ in RA and
$0.18''\pm0.02''$ in Dec, where the quoted uncertainties are the
standard deviations of the offsets for the seven objects.  As an
external test of our boresight correction, we note that our optical
position for PSR~J1740-5340 agrees with that of \citet{Bassa04} to
$0.01''$ in RA and $0.03''$ in Dec.  They followed a similar
astrometric procedure to what we used, with WFPC2 rather than ACS/WFC
imaging.

We searched for optical counterparts to the \chandra\ sources by
overlaying the \chandra\ error circles on the \hst\ stacked images,
with the boresight correction applied to the \chandra\ source
positions from \citet{Bogdanov10}.  The 95\% confidence error circle sizes
were computed following \citet{Hong05}.  Since the uncertainty in the
optical positions $(\lesssim 0.1\arcsec$) was small compared with the
size of the X-ray error circle radii ($\sim 0.5\arcsec$), we neglected
the contribution of the former to the total positional uncertainty.

A preliminary proper-motion analysis was carried out in order to do a
first-pass cluster-field separation.  The present dataset was used for
the reference epoch, with the stellar positions averaged over the 10
orbits.  Two different datasets were considered for the second epoch,
GO-7335 (obtained with the WFPC2 in 1999) and GO-10775 (obtained with
the ACS/WFC in 2006).   The former provides a longer time baseline,
while the later provides more depth and spatial coverage.  To compute
proper motions, we measured stars in each early-epoch exposure image,
corrected these positions for distortion, then transformed these
positions into the reference frame using linear transformations based
on only cluster members.  By using only members to define the
transformations, we automatically get displacements (and hence motions)
with respect to the cluster's systemic motion.

\begin{figure}
\includegraphics[width=3.4in]{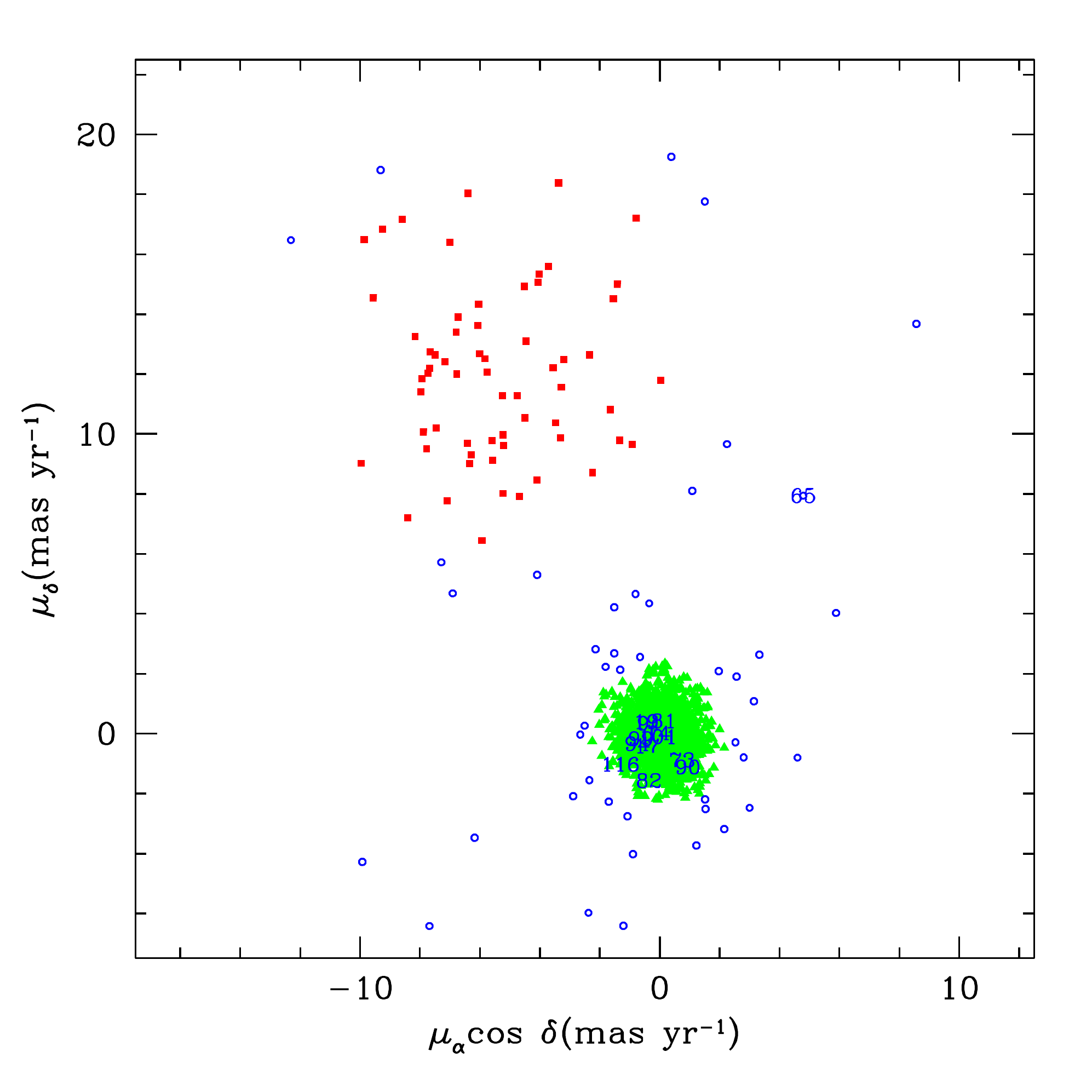}
\figcaption{Proper-motion components for those stars in the magnitude
  range $18 \le R < 19$ for which a proper-motion measurement was
  obtained.  The 2006 dataset was used for the second epoch for this
  plot.  The zeropoint corresponds, by design, to the systemic cluster
  motion.  Stars that are clearly in the cluster distribution are
  plotted as green triangles and fall in a tight clump about the
  origin, stars in the field distribution are plotted as red squares,
  and stars with an unknown status are plotted as blue circles.  The
  14 candidate identifications for stars in this magnitude range are
  indicated by blue numbers.  Note that source U65 belongs to the
  unknown group and U63 does not have a measured proper motion, while
  the other sources in this magnitude range are consistent with
  cluster membership.  A reasonable interpretation is that U65 is a
  relatively nearby active binary that is superposed on the cluster;
  see Fig.~\ref{f:CMD_AB} for the photometric status of this star.
  \label{f:PM}}
\end{figure}

The cluster-field separation that can be achieved for moderately
bright stars is illustrated in Fig.~\ref{f:PM}, which shows the
proper-motion distribution for stars with $18 \le R < 19$.  Cluster
and field regions are defined by a magnitude-dependent radius about
the centroids of the apparent cluster and field clumps.  Stars that do
not reside in either of these regions are considered to have an
unknown proper-motion membership status.  Since the proper-motion
error increases with magnitude, this proper-motion separation becomes
less effective with increasing magnitude.  Eventually the cluster and
field regions of the proper-motion plane show a strong overlap.  It
was generally possible to obtain an approximate discrimination between
cluster and field stars to a limiting magnitude of $R \approx 24$.  In
a few cases where the 2006-second-epoch results were resulted in an
unknown proper-motion membership status, the 1999-second-epoch results
were definitive.

\bigskip

\section{Results}

\subsection{\chandra\ Source Identification \label{source_ID}}

Optical CMDs provide a powerful means of selecting likely
\chandra\ source counterparts and investigating their properties.  For
each of the 79 \chandra\ sources within the half-mass radius, we
checked the locations of all objects within the X-ray error circle in
the CMDs shown in Figures~\ref{f:CMD_CV} and \ref{f:CMD_AB}.  Objects
that fell on the main sequence were considered to be unlikely
counterparts, given the relatively low X-ray to optical flux ratio,
$f_X/f_{\rm opt}$, of most MS stars, in contrast to the ranges for
chromospherically active binaries and cataclysmic variables.
Table~\ref{t:counterparts} summarizes the result of this counterpart
search.  Plausible identifications were obtained for nearly all
sources and are listed in this table, along with previous source
designations, inferred object types, proper-motion information, and
photometry.  The only sources for which the error circle was empty are
the quiescent low-mass X-ray binary (qLMXB) U24 \citep{Grindlay01},
U104, U106, U111, U113, U119, U121, U122, and U123\@.  In the case of
U24, there is a small ``blip'' near the center of the error circle in
the stacked $R$ image, but it likely represents the combination of
Airy ring artifacts from two bright nearby stars.  Several other
sources, U5, U16, U77, and U84, lie outside of the $R$ field and thus
are difficult to classify, from these data, due to the lack of
\br\ and \hr\ color measurements.  However, the U5 and U77
counterparts, V30 and V36 respectively, have been shown to be variable
using ground-based photometry \citep{Kaluzny06}.\footnote{Prefix V
  objects are variable stars from the ground-based studies of
  \citet{Kaluzny03} and \citet{Kaluzny06}.}  \citet{Kaluzny06}
classified V30 as an eclipsing binary that lies to the far red of the
main sequence (suggestive of it being a non-member).
\citet{Rozyczka10} have recently classified V36 as a single-line
spectropsopic binary.  We have classified both of these stars as
uncertain active binaries (ABs).  U5 and U84 have very red $B\!-\!\ha$
colors, suggesting that their \br\ colors are likely very red as well,
which supports an AB identification.  Table~\ref{t:counterparts} also
gives a summary of the results of the proper-motion analysis, which
indicates whether the object is consistent with being a cluster
member, a field member, or neither, based on a comparison of its
proper motion to the proper-motion distribution for stars of similar
magnitude.

\subsection{Source Types \label{source_types}}

Based on the location of the proposed counterparts in the CMDs (or, in
a few cases, based on ground-based detection of variability), we
primarily assigned types of cataclysmic variable (CV) and
chromospherically active binary (AB)\@.  CVs were defined as being
significantly to the blue of the MS and having significant
\ha\ excesses (either relative to the MS or to the white dwarf
sequence).  ABs were defined as lying within $\sim0.75$~mag above the
MS (and thus within $\sim0.2$~mag to the red of the MS) and having
small \ha\ excesses ($\la 0.1$~mag), based on the previous analysis of
ABs in NGC~6397 by \citet{Taylor01}.  In three cases where the only
object in the error circle was an apparent MS star, U41, U91, and
U112, we note its presence in Table~\ref{t:counterparts}.  Similarly,
an apparent MSTO star is the only object present in the error circle
for source U118\@.  We note that an AB with a low mass ratio and weak
lines could look like a MS star in both CMDs.

Two objects, U28 and U108, were classified as background galaxies
based on the extended appearance of their images.  U28 resembles an
edge-on spiral, while U108 has a more complex structure suggestive of
interacting galaxies.  The high X-ray to optical flux ratio of U28
(see \S\ref{flux_ratio}) suggests that it is an active galactic
nucleus (AGN)\@.  While the flux ratio of U108 was not determined, due
to the complex nature of the optical image, it is clearly much lower
than that of U28\@.  One possibility is that we are detecting X-ray
emission from U108 that is produced by a galaxy collision.  

Figure~\ref{f:CMD_CV} shows the location of the CV candidates in the
CMDs.  There is a suggestion of an evolutionary sequence from young,
bright CVs to old, faint ones.  We return to this point in
\S\ref{summary}.  The six brightest CVs mostly lie about 0.2 -- 0.8
mag to the blue of the MS\@. The optical emission of the systems
appears to be dominated by the secondary in the $R$ band, with a
larger contribution from the disk in the $B$ band. The fairly high
$R$-band flux indicates that the secondaries are relatively massive,
$\sim 0.5-0.7~\msun$, as inferred from the isochrones of
\citet{Baraffe97}.  All of the bright CVs have substantial
\ha\ excesses relative to the MS; these excesses generally increase
with magnitude.

The CV candidates 1--8 are numbered according to the scheme used in
previous papers, as listed by \citet{Grindlay01}.  A prefix CV is used
to designate each CV candidate.  \citet{Grindlay01} identify U28 as
a CV; however we find that this Chandra source appears to be a galaxy.
Here we use the identification by \citet{Taylor02} of U60 as CV9\@.
CVs 10--15 are newly identified in the present study;
\citet{Grindlay06} assigned numbers for CVs 10 and 11.  We have
numbered CVs 12--15 based on optical luminosity within this group.
The translation between \chandra\ source numbers and CV numbers is
given in Table~\ref{t:counterparts}.

There is a two magnitude gap between the bright CVs and the fainter
CVs.  Just below this gap are two possible transitional objects, U7
and U61, which lie between the white dwarf sequence and the main
sequence.  We note that the proper-motion data produce an unknown
membership status for U61, while U7 registers as consistent with
membership.  This leaves open the possibility that U61 is a background
AGN with an emission line that is shifted into the \ha\ window.
However, the image of the U61 counterpart does not show obvious
evidence of extension.  The fainter CV candidates, below U7 and U61,
lie in the vicinity of the WD cooling sequence.  The optical fluxes
for the faint CV candidates are clearly dominated by the contribution
of the WD\@.  The \hr\ indices for these faint CVs nearly all lie at
least 0.5 mag to the \ha\ excess side of the WD clump, which is
centered at about $\hr=3.25$. This suggests that the faint CVs have a
strong \ha-emission core (due to an accretion disk) within the broad
absorption lines of the WD continuum. This inferred spectroscopic
property of the faint CVs appears to be generally consistent with
those of the WZ Sge class of evolved cataclysmic variable
\citep{Schwarz04}.  Another possibility is that some of these objects
are AM~CVn type double-degenerate systems, in which a low-mass He~WD
donor feeds a much more massive carbon-oxygen WD\@.  Such systems show
no evidence of hydrogen lines in their spectra and instead have He~I
and sometimes He~II lines, which are typically in emission
\citep{AndersonS05,AndersonS08}.  AM~CVn stars could have evolved into
contact from the known He~WDs in NGC~6397, which have been shown to
have (detached) heavy CO WD companions \citep{Strickler09}.  Such an
object should lie close to the main sequence in the $(\hr,R)$ CMD\@.
Possible candidate AM~CVn stars among the CV candidates reported here
are U25, U31, U60, U80, and U83.

We note that most of the WD-like CV candidates cluster near a
magnitude of $R\sim23.5$, which corresponds to a $M_R \sim 11.2$,
using the distance modulus determined by \citet{Strickler09}.  We note
that \citet{Gaensicke09} have reported a sharp peak (``spike'') near
the period minimum ($P\approx80\!-\!86~{\rm min}$) in the period
distribution of CVs selected from the Sloan Digital Sky Survey
(SDSS)\@.  The average absolute magnitude for the CVs in the spike is
$\langle M_g \rangle = 11.6\pm0.7$.  Allowing for the difference in
photometric band, this is reasonably close to the characteristic
magnitude of the faint CVs in our study.  This suggests that these
faint CVs may well belong to the spike population, i.e.\ have periods
in the vicinity of the period minimum.  This could be tested by
determining orbital periods for the faint CVs from suitable
time-resolved photometry.  \citet{Gaensicke09} find that of 33 CVs
with orbital periods less than 86~min, 20 have spectra that are
WD-dominated.  In our sample, it appears that all seven faint CVs are
WD-dominated.  Given the small sample size, it is not clear whether
this represents a significant difference between faint field and faint
cluster CVs.

Figure~\ref{f:CMD_AB} shows the location of the objects identified as
AB stars in the CMDs\@.  We note that these objects sometimes lie near
the edge of the MS in either the right or left panel, but deviate by a
larger amount in the other panel.  Of the 42 AB counterparts listed in
Table~\ref{t:counterparts}, 26 are newly identified in our study.  Overall,
the ABs appear to form a relatively homogeneous binary sequence
alongside the MS, presumably mostly differing in mass.
\citet{Taylor01} have argued that these stars are likely BY Draconis
stars and represent a hard binary population in NGC~6397.  In a large
study of photometric variability in 47~Tuc, \citet{Albrow01} found a
substantial population of BY Draconis binaries, as well as W UMa and
other contact binaries, and eclipsing binaries.

In addition to the likely BY Dra population, there are a number of
counterparts that show a significantly different distribution in the
CMD\@.  These include the ``red straggler'' counterpart to U12 (the
millisecond pulsar) and the similar counterpart to U18, the
blue-straggler counterpart to U70, a group of five stars with very red
colors, and a group of five stars with strong \ha\ excesses.  There
are two stars in common between these latter two groups.

The counterparts to U12 and U18 both have proper motions consistent
with cluster membership.  Given their similar optical and X-ray
properties, it appears likely that U18 is a MSP \citep{Bogdanov10}.
The counterpart to U70, which registers as a proper-motion cluster
member, has been identified as variable star V20 by \citet{Kaluzny03}.
\citet{Rozyczka10} find that it is a double-line spectroscopic binary
with a period of $0.86\,{\rm d}$ and a mass ratio of $q\approx0.2$.
The He~WD candidate PC-5 \citep{Taylor01}, which has been previously
suggested as a possible counterpart to U70, lies well outside of the
current error circle.

The counterparts to U42, U63, U65, U86, and U92 differ strikingly from
the typical ABs, lying up to 1 mag to the red of the MS, and thus well
above it.  Of these, the counterparts to U42 and U92 have been
previously identified as optically variable stars \citep{Kaluzny06},
as confirmed here (see \S\ref{variability} and
Fig.~\ref{f:variability}); the counterparts to U63, U65, and U86 are
new identifications.  The U92 counterpart (V7) is classified as a W UMa
system by \citet{Kaluzny03}, while the U42 counterpart (V26) is
classified as an irregular variable by \citet{Kaluzny06}.  U92 is
consistent with the cluster proper-motion distribution.  The
proper-motion information available for U42, U65, and U86 produce an
unknown membership status.  These stars have proper motions that are
well outside of the cluster distribution but they are also
inconsistent with the field distribution.  There is no proper-motion
information for U63.  One reasonable possibility is that these stars
are foreground active binaries superposed on the cluster, as discussed
by \citet{Kaluzny06} in the case of U5 (V30).  In this interpretation,
the foreground binaries contain K/M type dwarfs with distance moduli
that put them well above the cluster main sequence in the $(\br,R)$
CMD\@.

Figure~\ref{f:CMD_AB} also shows several AB candidates that have
larger than typical \ha\ excesses; these include the counterparts to
U62, U65, U66, U86, and U120.  There is no proper-motion information
available for U62\@.  The proper-motion information for U65, U86, and
U120 produces an unknown membership status.  U66 is consistent with
the field proper-motion distribution.  Thus, the proper-motion
information does not offer much assistance in determining whether
these stars are likely cluster members.

\citet{Rozyczka10} have drawn attention to two of the variable stars
that we have identified as AB counterparts, V17 and V36, which
correspond to the sources U76 and U77\@.  These similar objects both
lie close to the main-sequence turnoff in the CMD presented by
\citet{Rozyczka10}.  Both objects are single-line spectroscopic
binaries, in which the unseen primary is inferred to have a mass in
excess of 1~\msun.  They suggest that these two objects are members of
a class of dormant degenerate binaries which may have much larger peak
X-ray luminosities.

\subsection{X-ray to Optical Flux Ratios \label{flux_ratio}}

\begin{figure}
\epsscale{1.2}
\includegraphics[width=3.4in]{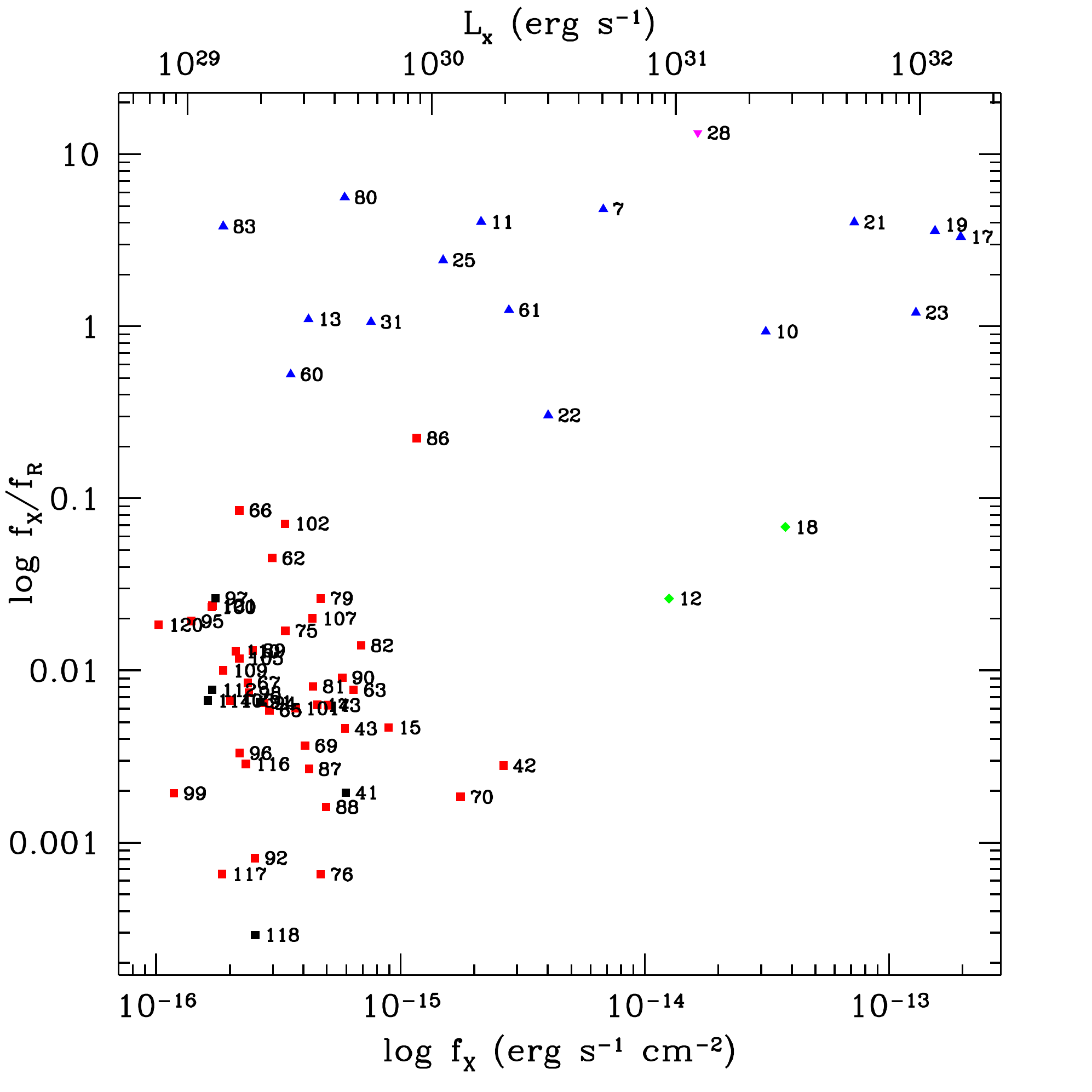}
\figcaption{X-ray to optical $R$-band flux ratio vs.\ X-ray flux
  (0.5--6 keV) for CVs (blue triangles), ABs (red squares), MSPs
  (green diamonds), unidentified objects and main sequence stars
  (black squares), and an AGN (inverted magenta triangle).  The
  $R$-band flux is taken as a measure of the optical flux. The upper
  axis gives the equivalent X-ray luminosity assuming that all objects
  are at the distance of the cluster.  Note that the ABs lie entirely
  below the CVs.  
\label{f:fx_fopt}}
\end{figure}

As an additional aid to the classification of sources, we have
examined the X-ray to optical flux ratio, $f_X/f_{\rm opt}$, where we
take $f_X (0.5\!-\!6\,{\rm keV})$ from \citet{Bogdanov10} and set
$f_{\rm opt} = f_R = 1.52\times10^{-0.4R-6} $.  The latter conversion
factor is computed from the \hst\ flux calibration constants.  The
X-ray count to flux transformation assumes a power-law spectrum with a
photon index of $\Gamma = 2.5$ \citep{Bogdanov10}.  The resulting flux
ratio is plotted versus $f_X$ in Fig.~\ref{f:fx_fopt}.  The ratio is
observed to be higher for accretional sources, such as CVs, LMXBs, and
AGNs, than for chromospherically active binaries \citep[see e.g.][who
  studied the \chandra\ source distribution in M4]{Bassa04b}.
Consistent with this, Fig.~\ref{f:fx_fopt} shows that the CV and AB
candidates form two distinct groups with no overlap in flux ratio.
The median flux ratio is about 500 times larger for the CVs than for
the ABs.  Within each group there is a broad range of the flux ratio,
about a factor of 40 for the CVs and a factor of 100 for the ABs.
There is no apparent dependence of the flux ratio on $f_X$ over a
range of more than three decades in $f_X$ for the CVs and nearly two
decades for the ABs.  The apparent AGN, U28, has the highest flux
ratio of all objects.  The known millisecond pulsar, U12, has a higher
flux ratio than most of the ABs and is about 40$\times$ brighter in
the X-ray than a typical AB\@.  The object U18, which has a similar
optical counterpart to that of U12, has even higher values of $f_X$
and $f_X/f_{\rm opt}$.  This supports the interpretation of U18 as a
MSP not yet detected at radio wavelengths \citep{Bogdanov10}.

Interestingly, the X-ray flux of CV5 (U22) is considerably lower in
the 2007 observations than in the previous 2002 observations; the flux
differs more than an order of magnitude.  This gives CV5 the lowest
$f_X/f_{\rm opt}$ value of all the CVs when the 2007 X-ray flux is
used together with the 2004-05 optical flux.  Since the X-ray and
optical fluxes were not measured contemporaneously, it is possible
that CV5 typically has a flux ratio that is more consistent with the
other bright CVs.  We note that the X-ray flux of a bright CV
typically drops during outburst \citep[e.g.][]{Wheatley03}, suggesting
that CV5 may have been in outburst in 2007.  The objects U62, U66, and
U86, which lie among the ABs with the four highest flux ratios, all
have significantly larger \ha\ excesses than the typical AB\@.  It is
not surprising that two measures of chromospheric activity, X-ray flux
and \ha\ flux, are correlated for these stars.

\subsection{Variability} \label{variability}

\begin{figure}
\epsscale{1.2}
\includegraphics[width=3.4in]{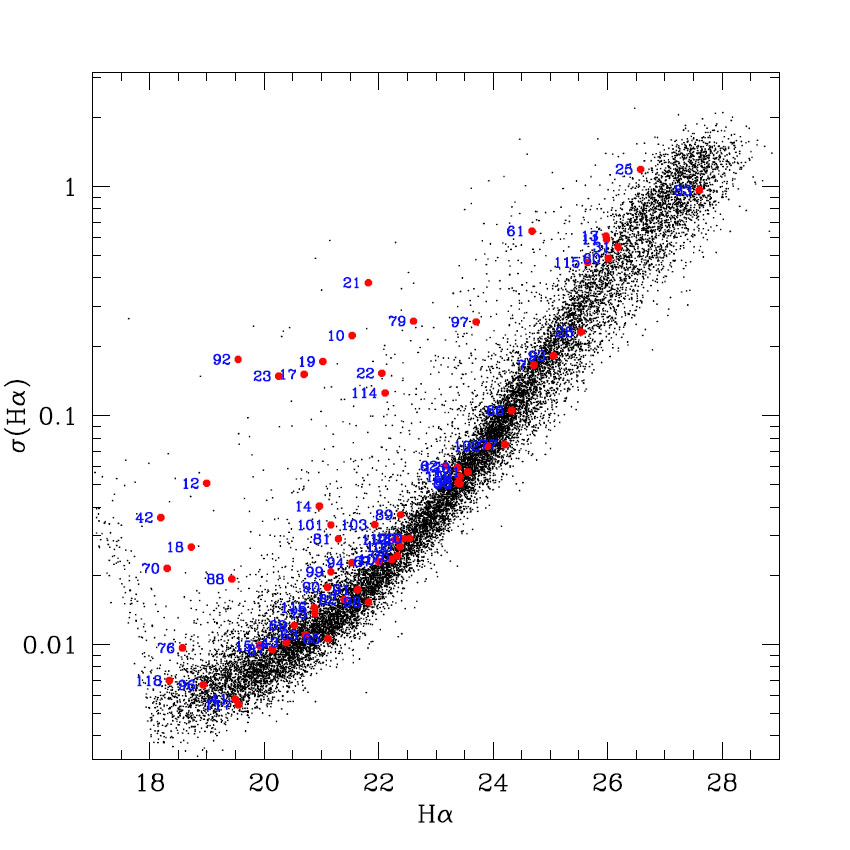}
\figcaption{\ha\ variability versus \ha\ magnitude.  The ordinate is
  the dispersion of the up to 40 \ha\ measurements for each star.  A
  3$\sigma$ clip has been used to filter out photometric problems.
  The set of stars that registers as variable, i.e.\ lies above the
  main distribution, is mostly independent of whether or not sigma
  clipping is applied.  Sources with at least nine measured magnitudes
  are plotted.  
\label{f:variability}}
\end{figure}

Since our dataset provides a 40-exposure time sequence of
\ha\ exposures, with four exposures per orbit, it was possible to
investigate optical variability.  The time sequence samples time
scales shorter than about one hour (the visibility period per
\hst\ orbit) and also time scales from one month to one year.  The
analysis of variability is complicated by the presence of outliers in
the time sequences, some of which represent photometric anomalies.  We
investigated several measures of variability and adopted the RMS
deviation about the mean of the \ha\ magnitude measurements, computed
using an iterative 3$\sigma$ clip to reduce the impact of outliers.
We plot $\sigma(\ha)$ versus mean \ha\ magnitude in
Figure~\ref{f:variability}.  This procedure results in a measure of
variability that is most sensitive to orbital variability of binary
systems, rather than large-amplitude fluctuations of CVs, given the
outlier filtering.  Thus, we also investigated the total range of the
\ha\ magnitudes for each object and constructed light curves for each
of the CV candidates in order to test for outburst behavior.

As can be seen in Fig.~\ref{f:variability}, most stars fall on a
well-defined sequence of increasing $\sigma(\ha)$ with increasing
magnitude.  Stars that lie significantly above this sequence generally
show evidence for variability, although some of this scatter is likely
due to photometric issues.  We note that the mean trend of the
variability index $\sigma(\ha)$ with mean $\ha$ magnitude differs for
stars brighter than $\ha \approx 18$, for which $\sigma(\ha)$
increases from a minimum of about 0.003 mag to about 0.04 mag for the
brightest stars plotted.  This behavior for bright stars is due to
saturation issues that increase in size with decreasing magnitude.

We have plotted the positions of the previous and new Chandra source
counterparts in Fig.~\ref{f:variability}.  Many of these counterparts
show significant variability as measured by the $\sigma$ of the \ha\
time series.  This group includes all of the bright CVs 1--6: sources
U10, U17, U19, U21, U22, and U23.  Of the newly identified faint CV
counterparts, only U61 registers as clearly variable by this measure
and U25 shows a possible slight variability.  While the other
faint CV candidates have $\sigma(\ha)$ values that are comparable in
size to those of the bright CVs, these values are not significantly
larger than those of other stars of similar magnitude.  Thus, the
failure to detect statistically significant variability for the other
faint CV counterparts may well be due to the decreasing sensitivity of
the variability test with increasing magnitude.

The total magnitude range for each of the six bright CVs was about
0.6--1.3~mag, which is less than the 2--5~mag amplitude for a dwarf
nova outburst \citep{Warner95}.  Source U21 (CV4) had the largest
amplitude (1.3 mag) of these bright CVs, with a light curve that a
high phase of about 50 days duration.  Source U61 (CV12) displayed a
total range of 2.3 mag.  The faintest magnitude measurements, which
contribute significantly to this large range, have a high degree of
uncertainty.  The light curve shows a clearly differentiated low state
in the first half of the year of observation and a high state during
the second half.  The mean offset between these two states is about 1
mag and thus also falls short of a dwarf nova outburst.

A number of the candidate AB counterparts also show evidence of
variability, including six newly identified ABs: sources U79, U97, and
U114, which show strong variability at the level of the six bright
CVs; sources U88 and U101, which show moderate variability; and source
U103, which shows slight evidence of variability.  Five previously
identified AB counterparts, U14, U18, U42, U70, and U92, show clear
variability, and U76 and U81 show weaker evidence of variability.  We
note that U18 has a similar level of variability to the MSP
counterpart U12, again suggesting that these are similar objects.
 
We carried out a period-folding light curve analysis for the bright
CVs 1--6 and the MSP U12\@.  Given the very unequal spacing of the
data, it is difficult to determine a period ab initio, due to the high
degree of aliasing.  However, we were able to confirm the previously
determined orbital periods obtained by \citet{Kaluzny03} and
\citet{Taylor02} for CVs 1 and 6, as well as that of U12
\citep{DAmico01}. We also see evidence for additional long-term
variability for CVs 1 and 6, at the level of a few tenths of a
magnitude.

\subsection{Spatial Distribution}

\begin{figure}
\epsscale{1.2}
\includegraphics[width=3.4in]{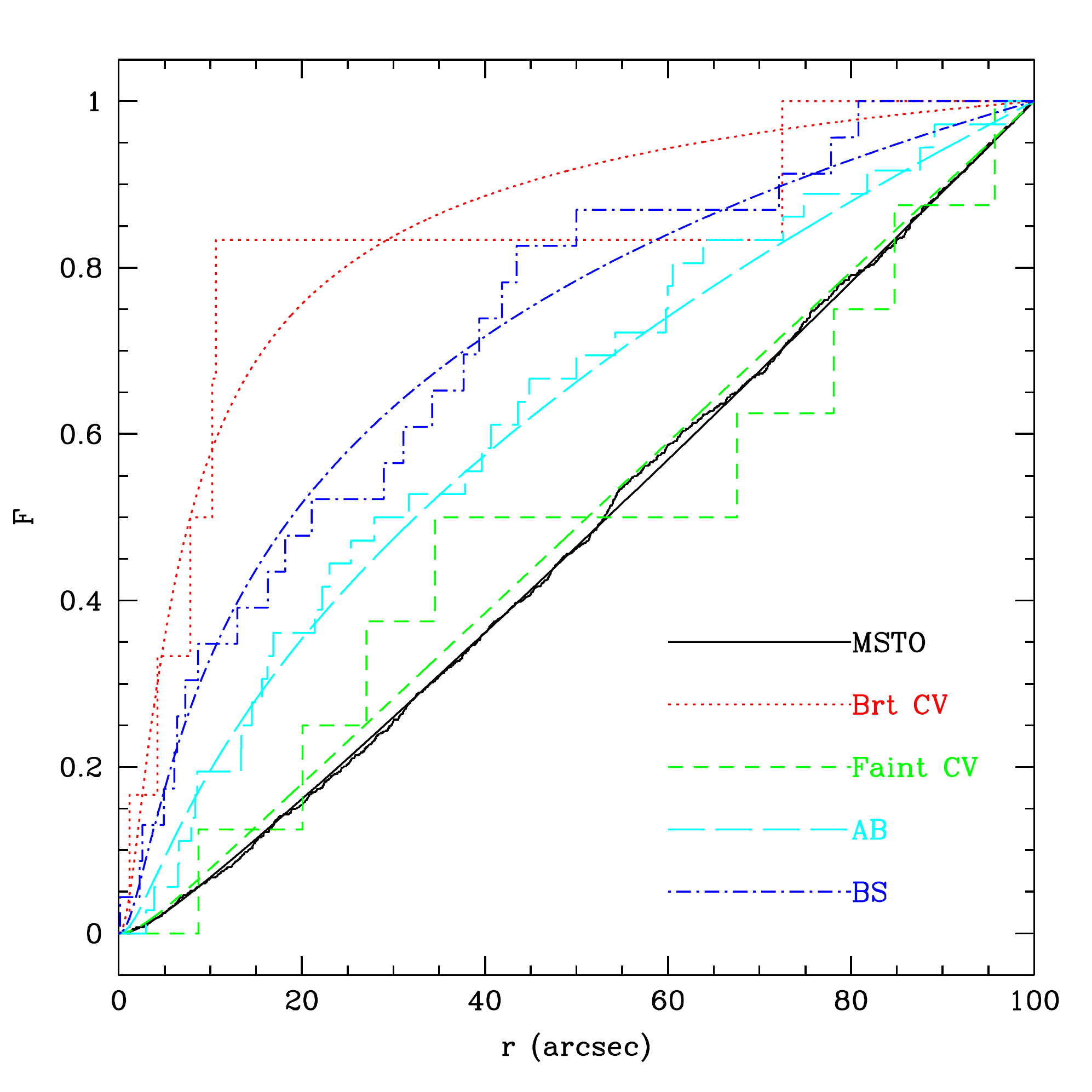}
\figcaption{Cumulative radial distributions for selected stellar
  groups.  For each group the actual distribution (stair-step curve)
  and maximum likelihood fit (smooth curve) are shown.  Note that the
  bright CVs show the highest degree of central concentration,
  followed by the blue stragglers.  The distribution of faint CVs is
  statistically consistent with the MSTO stars.  Fitting information
  for these stellar groups is given in Table~\ref{t:Cored_PL_fits}.  
\label{f:radial}}
\end{figure}

We determined the cluster center by iterative centroiding in a
20\arcsec\ radius aperture using a sample of MS stars with magnitudes
in the range $16 \le R < 22$.  The resulting center of $\alpha =
17^{\rm h}~40^{\rm m}~42.17^{\rm s},~ \delta = -53^\circ~40'~28.6''$
lies within about 1\arcsec\ of CV2 (U19)\@.  Experimentation with the
centroiding aperture size and the stellar sample definition indicates
that the center position is uncertain by about 1\arcsec.  We then
determined the cumulative radial distributions of a number of stellar
groups out to a radius of 100\arcsec, which is the approximate size of
the central surface density cusp of NGC~6397 \citep{Lugger95}.
Figure~\ref{f:radial} shows the distribution functions for the stellar
groups listed in the figure legend.  As has been previously noted, the
bright CVs in NGC~6397 show a strong central concentration relative to
other stellar groups, including the fainter CVs.  With the exception
of CV6, the five other bright CVs lie within 11\arcsec\ ($2 r_c$) of
the cluster center.  In comparison, with the exception of U25, the
faint CVs lie at least 20\arcsec\ ($4 r_c$) from the cluster center.
We carried out Kolmogorov-Smirnov intercomparisons of the stellar
samples defined in Fig.~\ref{f:radial}.  The comparison of each group
to a MSTO sample, defined by $16 \le R < 17$, is given in
Table~\ref{t:Cored_PL_fits}, where the probability, $p$, of the two
samples being drawn from the same parent distribution is listed. The
bright CV and blue straggler (BS) samples both differ very
significantly from the MSTO sample ($p<0.1\%$), while the overall CV
and AB samples differ from the MSTO sample at the 4\% level and 1\%
level, respectively.  The bright and faint CV samples differ from each
other at a significance level of 3\% and the bright CV and AB samples
differ at a level of 2\%.

\begin{deluxetable*}{lrcccccr}
\tabletypesize{\normalsize}
\tablenum{2}
\tablecolumns{8}
\tablecaption{\textbf{Cored Power-Law Model Fits to 100 Arcsec} 
\label{t:Cored_PL_fits}}
\tablehead{
\colhead{Sample} & 
\colhead{N\tablenotemark{a}} &
\colhead{$q$} & 
\colhead{$r_c~(\arcsec)$} &
\colhead{$\alpha$} &
\colhead{$m~(\msun)$} &
\colhead{$\sigma$\tablenotemark{b}} &
\colhead{K-S prob\tablenotemark{c}}
} 
\startdata
MSTO       & 1111 &   1.0            & $5.5 \pm 3.4$ & $-0.93 \pm 0.05$ & $0.80 \pm 0.05$ & \nodata & \nodata \\
CV         &   14 &  $1.42 \pm 0.16$ & $3.3 \pm 2.1$ & $-1.74 \pm 0.32$ & $1.14 \pm 0.14$ & 2.3     & 4.0\%  \\
bright CV  &    6 &  $1.83 \pm 0.26$ & $2.5 \pm 1.6$ & $-2.53 \pm 0.51$ & $1.46 \pm 0.22$ & 2.9     & 0.07\% \\
faint CV   &    8 &  $1.04 \pm 0.24$ & $5.1 \pm 3.8$ & $-1.01 \pm 0.47$ & $0.83 \pm 0.20$ & 0.2     & 89\%   \\
AB         &   36 &  $1.32 \pm 0.08$ & $3.6 \pm 2.2$ & $-1.54 \pm 0.17$ & $1.06 \pm 0.08$ & 2.6     & 1.1\%  \\
BS         &   23 &  $1.52 \pm 0.10$ & $3.0 \pm 1.9$ & $-1.93 \pm 0.21$ & $1.22 \pm 0.10$ & 3.7     & 0.03\% \\
\enddata
%
\tablenotetext{a}{Size of sample within 100\arcsec\ of cluster center}
\tablenotetext{b}{Significance of mass excess above MSTO mass}
\tablenotetext{c}{K-S probability of consistency with MSTO group}
\end{deluxetable*}

In order to further investigate the spatial distribution of the
\chandra\ sources in NGC~6397 and the implications for object masses,
we carried out maximum likelihood fits of what has been termed a
``generalized King model'' to the surface density distributions.  This
model can be described more generally as a ``cored power law,'' in
that it has a smooth transition from an inner core region to an outer
power law.  It takes the form,
\begin{equation}
\label{eqn:Cored_PL} 
S(r) = S_0 \left[1 + \left({r \over r_0}\right)^2 \right]^{\alpha/2},
\end{equation}
with the core radius $r_c$ related to the scale parameter $r_0$ by,
\begin{equation}
r_c = \left(2^{-2/\alpha} -1 \right)^{1/2} r_0\,.
\end{equation}
An ``analytic King model,'' which provides a good fit to the surface
density profile of the turnoff-mass stellar group for a normal-core
cluster \citep[e.g.\ 47~Tuc;][]{Heinke05}, corresponds to $\alpha =
-2$ and $r_c = r_0$.  Since NGC~6397 has a central-cusp structure, due
to its core-collapsed status, we expect that the turnoff-mass group
will have a flatter power-law slope than $\alpha = -2$ \citep{Cohn85}.
We performed a maximum-likelihood fit of Eqn.~\ref{eqn:Cored_PL} to a
turnoff-mass group, defined by $16 \le R < 17$ and radial offset
within $100\arcsec$ of the cluster center, to determine the best-fit
values of $\alpha$ and $r_c$.  We used nonlinear optimization to
maximize the likelihood and bootstrap resampling to estimate the
parameter uncertainties, as discussed by \citet{Grindlay02}.  The
resulting best-fit parameter values are $\alpha = -0.93 \pm 0.05$ and
$r_c = 5.5\arcsec \pm 3.4\arcsec$.

By fitting Eqn.~\ref{eqn:Cored_PL} to the surface density
distributions of individual groups such as the CVs and ABs, the
characteristic masses of these objects can be estimated, as described
by \citet{Heinke05} and \citet{Lugger07}.  In order to do this, we
adopt the approximation here that stellar groups with masses exceeding
the turnoff mass are in approximate thermal equilibrium with the
turnoff mass group and that the velocity dispersion of each mass group
is approximately constant with radius within the central cusp region.
This approach is motivated by Fokker-Planck simulations of
post-collapse core oscillations which indicate that a cluster core
spends a majority of the time near a maximally expanded state in which
the more massive stellar groups are close to thermal equilibrium
\citep{Murphy90}.  More recent Monte-Carlo and N-body simulations
suggest that the presence of primordial binaries may delay the
gravothermal oscillation phase to well beyond a Hubble time and that
even clusters with an observationally core-collapsed designation such
as NGC~6397 may still be in the binary burning phase
\citep{Fregeau08}.  In any case, the assumption of thermal equilibrium
among the most massive components in the central region of an evolved
cluster is a useful starting point.

In the thermal equilibrium approximation, the surface density profile
for a mass group with mass $m$ is given by Eqn.~\ref{eqn:Cored_PL} with
a slope parameter $\alpha$ related to the turnoff-mass slope
$\alpha_{\rm to}$ by
\begin{equation}
\alpha = q\, (\alpha_{\rm to} - 1) + 1
\end{equation}
where $q = m/m_{\rm to}$.  Thus, fitting this model involves just one
free parameter, the mass ratio $q$, as $S_0$ is determined by
normalization and $r_0$ and $\alpha_{\rm to}$ are determined by the
fit to the turnoff-mass group.

Table~\ref{t:Cored_PL_fits} gives the results of maximum-likelihood
fits of Eqn.~\ref{eqn:Cored_PL} to the turnoff mass stars, CVs, ABs,
and a comparison group of blue stragglers (BSs).  The cumulative
radial distributions for these groups and the model fits are shown in
Fig.~\ref{f:radial}.  As can be seen from the table, the $q$ values
for all of the groups exceed unity, indicating that the characteristic
masses exceed the turnoff mass.  For all cases other than the faint
CVs, the excesses are significant at the 2$\sigma$ or higher level.
The groups with the strongest evidence for a mass significantly above
the turnoff mass are the bright CVs ($2.9\sigma$) and the BSs
($3.7\sigma$).  We note that the results of this analysis agree with
the K-S comparison results given in the last column of the table.  The
inferred mass range for the bright CVs ($1.5 \pm 0.2\msun$) is
consistent with a system made up of a heavy white dwarf (e.g.\ $M_{\rm
  WD} \sim 0.9 \msun$) with a main sequence secondary of mass $\sim
0.6 \msun$.  The inferred mass range for the faint CVs ($0.8 \pm
0.2\msun$) is consistent with a somewhat less massive white dwarf
(e.g\ $\sim 0.7~\msun$) and a secondary mass that has been whittled
down to $\sim 0.1 \msun$.  In such a system, the optical flux would be
dominated by the white dwarf, as observed here.  We presume that the
white dwarf in this case has been heated by the accretion process,
since noninteracting heavy white dwarfs are too faint to detect due to
the degree of cooling that has occurred over their long lifetimes.  As
discussed in \S\ref{source_types}, an alternative interpretation is
that some of the CVs are AM~CVn systems, in which the secondary is a
He~WD\@.  In this case, the mass of the secondary would likely be
about 0.2--0.3~\msun \citep{Strickler09}.  It is interesting to
compare the white dwarf masses inferred here with those for field CVs.
\citet{Southworth09} have recently determined a mass of
$0.78\pm0.12~\msun$ for the WD in a field CV that lies above the
period gap.  \citet{Littlefair08} have determined the masses of the WDs in
seven CVs that lie below period gap, with a resulting mean mass and
standard deviation of $0.87\pm0.07~\msun$.  Thus, given the
uncertainty ranges in the CV masses inferred here, our results for
likely WD masses are generally consistent with these measured WD
masses.

\section{Summary} \label{summary}

We have shown that with moderately deep $B$, $R$, and \ha\
\hst-ACS/WFC imaging, it is possible to identify and classify nearly
all of the 79 currently detected \chandra\ sources within the
half-mass radius of NGC~6397.  Our classifications are based on CMD
location and are tested for consistency with expected ranges of the
X-ray to optical flux ratio.  In only a few cases was it not possible
to classify the source, generally due to a lack of coverage in one
filter.  The vast majority of the sources in the cluster can be
classified as either cataclysmic variables (a total of 15) or active
binaries (a total of 42).  The X-ray to optical flux ratio for the CVs
(0.2 -- 6) substantially exceeds that for most of the ABs
($0.0003-0.2$).  There is one previously known MSP and one potential
MSP; the latter classification is based on the similar photometric
properties of the two objects.

The proper-motion test for cluster membership produces an unknown
status for many of the \chandra\ source counterparts.  This is largely
a consequence of the short time baseline of 1.5 years between the two
epochs of ACS/WFC data used for determining proper motions.  While the
1999 WFPC2 observations give a longer baseline, fewer stars are
detected by the WFPC2\@.  Thus, the proper-motion part of this study
would benefit substantially from new \hst\ observations of NGC~6397
with either the ACS/WFC or the WFC3\@.  While it appears likely that
the vast majority of the proposed counterparts are cluster members,
proper-motion confirmation would be valuable.  It would be
particularly interesting to know how many of the atypical ABs are
cluster members.  While it is likely that most are foreground stars,
one of the very red ones has appears to be a member on the basis of
our preliminary proper-motion measurements.

The CV distribution in NGC 6397 has a bimodal character.  We detect
distinct bright and faint populations with only two possible
transition objects.  The optical emission of the six brightest CVs
appears to be dominated by a relatively massive secondary, while that
of the faint CVs appears to be dominated by the WD, with very little
contribution from a very low-mass secondary.  Alternatively, some of
the faint CVs may be AM~CVn systems, with a He~WD secondary.  The
distribution of CVs in optical luminosity and color is consistent with
expected CV evolution due to the effects of mass transfer and angular
momentum loss.  As CVs age and the secondary loses mass to the
primary, the orbit tightens due to magnetic braking and ultimately
gravitational wave radiation.  In this picture, the bright CVs 1--6
represent young, recently formed systems, while the faint CVs
represent old, highly evolved systems.  While CVs can be born at any
orbital period, depending on the nature of the secondary star, there
is a bias for CVs formed in exchange interactions in collapsed
cluster cores to have more massive secondaries and thus longer orbital
periods than newly formed field CVs.  This is both the result of mass
segregation in the cluster, which may invert the mass function in the
cluster core, and of the tendency for more massive stars to displace less
massive ones in exchange interactions.  Thus, while it is possible for
a faint CV to be a young system, we expect that the bulk of the faint
CVs evolved from an earlier bright state.

Given the likely ongoing formation of CVs by dynamical processes in
the central region of the cluster, it is plausible that we are
observing a roughly equilibrium CV population in NGC 6397.  However,
there remains a possibility that the current population of six bright
CVs represents the result of a recent core collapse event that
resulted in an enhanced production of CVs.  In this case, the bright
CVs would be overpopulous relative to the faint CVs.

The six brightest CVs in NGC~6397 have a much more centrally
concentrated spatial distribution than do either the fainter CVs or
the ABs.  Indeed, five of six bright CVs are located within about
2\,$r_c$ of the cluster center, while all but one of the fainter CVs
lie outside of 4\,$r_c$.  This suggests that CVs are preferentially
formed by dynamical processes within the core and surrounding region
and migrate to larger radii as they age and undergo repeated
scattering interactions that act to increase the typical size of their
orbits in the cluster potential.  Since repeated strong interactions
between compact binaries and other stars ultimately result in the
ejection of the binaries from clusters, the lifetime of a compact
binary in a cluster is extended if the amount of time it spends in the
central region of the cluster is reduced.  Thus, it is reasonable to
expect that long-lived CVs in globular clusters should generally be
found outside of the densest regions.  This is consistent with the
spatial distribution of the CVs in NGC~6397.

\acknowledgements{This work is supported by NASA grant HST-GO-10257A
  and NSF REU grant AST-0452975 to Indiana University, and NSERC grants
  to C.O.H\@.  S.B. is supported by a CIFAR Junior Fellowship.}





\clearpage

\LongTables
\newcommand{\nd}{\nodata}
\tabletypesize{\footnotesize}
\tabletypesize{\scriptsize}
\enlargethispage{0.25in}
\begin{landscape}
\begin{deluxetable}{llrlllllcccl}
\tablenum{1}
\tablecolumns{11}
\tablecaption{\textbf{Optical Counterpart Summary}\label{t:counterparts}}
\tablehead{
\colhead{Source\tablenotemark{a}} &
\colhead{RA, Dec (J2000)} &
\colhead{r ($'$)\tablenotemark{b}} &
\colhead{Previous/New IDs\tablenotemark{c}} &
\colhead{Detect?\tablenotemark{d}} &
\colhead{Type\tablenotemark{e}} &
\colhead{PM$_{06}$\tablenotemark{f}} &
\colhead{PM$_{99}$\tablenotemark{g}} &
\colhead{$R$} &
\colhead{$B$} & 
\colhead{$\ha$} &
\colhead{Notes}
}
\startdata
%
%
U5   &  17:40:54.531  $-$53:40:44.57   &  1.85    & V30             & 1         & AB?    & ?   & \nd  &  \nd    &  18.73  &  19.23 & outside of $R$ field \\  
U7   &  17:40:52.832  $-$53:41:21.77   &  1.81    & CV10$^*$        & 1         & CV     & c   & \nd  &  22.58  &  23.44  &  24.73 & \\
U10  &  17:40:48.978  $-$53:39:48.62   &  1.21    & CV6, V13        & 1         & CV     & c   & c    &  19.14  &  20.34  &  21.53 & \\
U11  &  17:40:45.781  $-$53:40:41.52   &  0.58    & CV7             & 1         & CV     & ?   & c    &  23.65  &  24.26  &  25.88 & \\ 
U12  &  17:40:44.621  $-$53:40:41.60   &  0.42    & BY-WF4-1, V16   & 1         & MSP    & c   & c    &  16.25  &  17.43  &  18.99 & \\
U13  &  17:40:44.084  $-$53:40:39.17   &  0.33    & CV8             & 1         & CV     & ?   & c    &  24.00  &  24.33  &  25.95 & \\
U14  &  17:40:43.328  $-$53:41:55.46   &  1.46    & V35             & 1         & AB     & c   & \nd  &  18.31  &  19.72  &  20.96 & \\
U15  &  17:40:42.910  $-$53:40:33.81   &  0.14    & BY-PC-2         & 1         & AB     & c   & \nd  &  17.25  &  18.53  &  19.93 & \\
U16  &  17:40:42.636  $-$53:42:15.24   &  1.78    & \nd             & 1         & ?      & ?   & \nd  &  \nd    &  \nd    &  25.83 & outside of $R$ and $B$ fields\\
U17  &  17:40:42.651  $-$53:40:19.30   &  0.17    & CV3, V33        & 1         & CV     & c   & c    &  18.52  &  19.10  &  20.67 & \\
U18  &  17:40:42.606  $-$53:40:27.62   &  0.07    & BY-PC-8, V31    & 1         & MSP    & c   & c    &  16.10  &  17.51  &  18.73 & \\ 
U19  &  17:40:42.306  $-$53:40:28.70   &  0.02    & CV2, V34        & 1         & CV     & c   & c    &  18.87  &  20.02  &  21.03 & \\
U21  &  17:40:41.830  $-$53:40:21.37   &  0.13    & CV4             & 1         & CV     & c   & c    &  19.82  &  20.94  &  21.82 & \\
U22  &  17:40:41.701  $-$53:40:29.00   &  0.07    & CV5             & 1         & CV     & c   & c    &  20.15  &  21.50  &  22.04 & \\
U23  &  17:40:41.597  $-$53:40:19.30   &  0.18    & CV1, V12        & 1         & CV     & c   & c    &  17.88  &  18.93  &  20.25 & \\
U24  &  17:40:41.468  $-$53:40:04.43   &  0.42    & \nd             & 0?        & qLMXB  & \nd & \nd  &  \nd    &  \nd    &  \nd   & hint of detection in $R$ only\\
U25  &  17:40:41.237  $-$53:40:25.79   &  0.15    & CV13$^*$        & 1         & CV?    & \nd & \nd  &  23.48  &  23.44  &  26.01 & uncertain photometry \\ 
U28  &  17:40:38.904  $-$53:39:51.09   &  0.79    & \nd             & 1         & AGN    & ?   & \nd  &  22.72  &  24.73  &  25.44 & \\
U31  &  17:40:34.202  $-$53:41:15.28   &  1.41    & CV11$^*$        & 1         & CV     & f   & \nd  &  23.32  &  23.93  &  26.07 & \\
U41  &  17:40:45.008  $-$53:39:55.21   &  0.70    & \nd             & 1         & MS     & c   & c    &  16.74  &  17.69  &  19.50 & only object in error circle \\
U42  &  17:40:43.059  $-$53:38:31.29   &  1.96    & V26             & 1         & AB     & ?   & \nd  &  15.52  &  17.47  &  18.20 & very red \\
U43  &  17:40:40.543  $-$53:40:22.79   &  0.26    & BY-PC-4         & 1         & AB     & c   & c    &  17.68  &  19.03  &  20.38 & \\
U60  &  17:40:47.807  $-$53:41:28.40   &  1.30    & CV9             & 1         & CV     & c   & c    &  23.38  &  23.63  &  26.04 & \\
U61  &  17:40:45.223  $-$53:40:28.60   &  0.45    & CV12$^*$        & 1         & CV?    & ?   & \nd  &  22.08  &  23.19  &  24.19 & uncertain photometry \\
U62  &  17:40:30.422  $-$53:39:17.47   &  2.11    & \nd             & 1         & AB?    & \nd & \nd  &  20.90  &  23.33  &  23.17 & uncertain photometry \\
U63  &  17:40:31.663  $-$53:38:46.36   &  2.31    & \nd             & 1         & AB?    & \nd & \nd  &  18.15  &  20.39  &  20.72 & very red, moderate \ha\ excess \\
U65  &  17:40:37.558  $-$53:39:17.85   &  1.36    & \nd             & 1         & AB?    & ?   & \nd  &  18.72  &  21.27  &  21.11 & very red, large \ha\ excess \\
U66  &  17:40:38.918  $-$53:38:49.80   &  1.71    & \nd             & 1         & AB?    & f   & \nd  &  21.93  &  24.56  &  24.29 & red, large \ha\ excess \\
U67  &  17:40:40.065  $-$53:40:16.59   &  0.37    & \nd             & 1         & AB     & c   & c    &  19.34  &  21.17  &  22.00 & \\
U68  &  17:40:40.730  $-$53:38:32.63   &  1.95    & \nd             & 1?        & ?      & \nd & \nd  &  \nd    &  \nd    &  \nd   & marginal detection in $R$ only \\
U69  &  17:40:40.867  $-$53:40:17.17   &  0.27    & BY-PC-5         & 1         & AB     & c   & c    &  17.84  &  19.03  &  20.52 & \\
U70  &  17:40:41.693  $-$53:40:33.33   &  0.11    & V20             & 1         & AB, BS & c   & c    &  15.50  &  16.21  &  18.30 & \\
U73  &  17:40:42.681  $-$53:39:28.73   &  1.00    & \nd             & 1         & AB     & c   & c    &  18.19  &  19.47  &  20.88 & \\
U75  &  17:40:43.666  $-$53:40:30.61   &  0.22    & \nd             & 1         & AB     & c   & \nd  &  19.71  &  21.78  &  22.32 & \\
U76  &  17:40:43.818  $-$53:41:16.32   &  0.83    & V17             & 1         & AB     & c   & c    &  15.81  &  16.81  &  18.57 & \\
U77  &  17:40:44.125  $-$53:42:11.48   &  1.74    & V36             & 1         & AB?    & \nd & \nd  &  \nd    &  \nd    &  24.21 & outside of $R$ field \\
U79  &  17:40:46.409  $-$53:40:04.05   &  0.75    & \nd             & 1         & AB     & ?   & \nd  &  19.82  &  21.98  &  22.41 & \\
U80  &  17:40:46.455  $-$53:41:56.50   &  1.60    & CV14$^*$        & 1         & CV?    & \nd & \nd  &  25.40  &  25.84  &  27.91 & weak \ha\ detection \\
U81  &  17:40:46.481  $-$53:41:15.44   &  1.01    & V14             & 1         & AB     & c   & c    &  18.62  &  20.27  &  21.29 & \\ 
U82  &  17:40:48.537  $-$53:39:39.53   &  1.25    & BY-WF2-1        & 1         & AB     & c   & \nd  &  18.72  &  20.29  &  21.39 & \\
U83  &  17:40:49.615  $-$53:40:43.02   &  1.13    & CV15$^*$        & 1         & CV?    & ?   & ?    &  26.22  &  27.18  &  28.56 & very faint \\
U84  &  17:40:54.807  $-$53:40:19.79   &  1.88    & \nd             & 1         & AB?    & ?   & ?    &  \nd    &  17.27  &  18.23 & very red, outside of $R$ field \\
U86  &  17:40:37.473  $-$53:41:47.24   &  1.48    & \nd             & 1         & AB?    & ?   & \nd  &  21.16  &  24.06  &  23.39 & very red, large \ha\ excess \\
U87  &  17:40:42.877  $-$53:40:26.45   &  0.11    & BY-PC-3         & 1         & AB     & c   & c    &  17.46  &  18.69  &  20.14 & \\
U88  &  17:40:42.863  $-$53:40:23.43   &  0.13    & BY-PC-1, V19    & 1         & AB     & c   & c    &  16.73  &  17.77  &  19.43 & \\
U89  &  17:40:43.613  $-$53:40:24.60   &  0.23    & \nd             & 1         & AB     & c   & c    &  19.76  &  21.82  &  22.38 & \\
U90  &  17:40:41.779  $-$53:40:14.42   &  0.24    & BY-PC-6         & 1         & AB     & c   & c    &  18.45  &  19.88  &  21.11 & \\
U91  &  17:40:42.430  $-$53:40:41.65   &  0.22    & \nd             & 1         & MS     & c   & \nd  &  18.94  &  20.44  &  21.63 & only object in error circle \\
U92  &  17:40:43.916  $-$53:40:35.39   &  0.28    & BY-WF4-2, V7    & 1         & AB?    & c   & c    &  16.72  &  18.26  &  19.53 & very red, slight \ha\ deficit \\
U93  &  17:40:42.393  $-$53:40:46.62   &  0.30    & \nd             & 1         & ?      & \nd & \nd  &  \nd    &  \nd    &  24.95 & uncertain photometry \\
U94  &  17:40:42.868  $-$53:40:49.07   &  0.36    & \nd             & 1         & AB     & c   & c    &  18.88  &  20.58  &  21.53 & \\
U95  &  17:40:40.320  $-$53:40:44.58   &  0.38    & \nd             & 1         & AB?    & c   & \nd  &  20.81  &  23.02  &  23.41 & MS color, \ha\ excess \\
U96  &  17:40:39.097  $-$53:40:23.09   &  0.47    & V24             & 1         & AB     & c   & \nd  &  18.40  &  20.15  &  21.42 & \\
U97  &  17:40:43.918  $-$53:40:05.90   &  0.46    & \nd             & 1         & ?      & c   & \nd  &  20.90  &  22.88  &  23.56 & blue, normal \ha\ \\
U98  &  17:40:40.994  $-$53:40:58.40   &  0.53    & \nd             & 1         & AB     & c   & \nd  &  19.18  &  21.09  &  21.82 & \\
U99  &  17:40:46.431  $-$53:40:30.40   &  0.63    & \nd             & 1         & AB     & c   & \nd  &  18.49  &  19.96  &  21.16 & \\
U100 &  17:40:38.201  $-$53:40:46.55   &  0.66    & \nd             & 1         & AB     & c   & \nd  &  20.81  &  23.14  &  23.40 & \\
U101 &  17:40:45.399  $-$53:41:01.40   &  0.73    & \nd             & 1         & AB?    & c   & c    &  18.48  &  19.94  &  21.19 & red, normal \ha\ \\
U102 &  17:40:38.845  $-$53:39:43.12   &  0.90    & \nd             & 1         & AB     & c   & \nd  &  21.27  &  23.80  &  23.87 & \\
U103 &  17:40:35.698  $-$53:40:12.56   &  1.00    & \nd             & 1         & AB     & c   & \nd  &  19.26  &  21.05  &  21.93 & \\
U104 &  17:40:43.124  $-$53:39:29.04   &  1.01    & \nd             & 0         & \nd    & \nd & \nd  &  \nd    &  \nd    &  \nd   & empty error circle \\
U105 &  17:40:36.521  $-$53:41:07.85   &  1.06    & \nd             & 1         & AB     & ?   & \nd  &  19.78  &  21.84  &  22.37 & \\
U106 &  17:40:43.737  $-$53:39:17.52   &  1.21    & \nd             & 0         & \nd    & \nd & \nd  &  \nd    &  \nd    &  \nd   & empty error circle \\
U107 &  17:40:34.115  $-$53:40:17.01   &  1.21    & \nd             & 1         & AB     & c   & \nd  &  19.61  &  21.63  &  22.23 & \\
U108 &  17:40:52.099  $-$53:39:48.25   &  1.62    & \nd             & 1         & GLX?   & \nd & \nd  &  \nd    &  \nd    &  \nd   & extended object \\
U109 &  17:40:52.728  $-$53:40:52.88   &  1.61    & \nd             & 1         & AB?    & c   & c    &  19.77  &  21.83  &  22.44 & red, normal \ha\ \\
U110 &  17:40:33.455  $-$53:39:16.83   &  1.76    & \nd             & 1         & AB     & c   & \nd  &  19.92  &  22.04  &  22.53 & \\
U111 &  17:40:29.845  $-$53:40:26.99   &  1.82    & \nd             & 0         & \nd    & \nd & \nd  &  \nd    &  \nd    &  \nd   & empty error circle \\
U112 &  17:40:50.374  $-$53:39:06.00   &  1.84    & \nd             & 1         & MS     & f   & \nd  &  19.60  &  21.25  &  22.29 & only object in error circle \\
U113 &  17:40:42.764  $-$53:40:20.76   &  0.16    & \nd             & 0         & \nd    & \nd & \nd  &  \nd    &  \nd    &  \nd   & empty error circle \\
U114 &  17:40:43.469  $-$53:40:34.34   &  0.21    & \nd             & 1         & ?      & c   & \nd  &  19.49  &  \nd    &  22.10  & uncertain photometry \\
U116 &  17:40:42.236  $-$53:40:19.97   &  0.18    & \nd             & 1         & AB     & c   & c    &  18.18  &  19.49  &  20.88 & \\
U117 &  17:40:42.153  $-$53:40:25.56   &  0.14    & \nd             & 1         & AB     & c   & c    &  16.82  &  17.84  &  19.55 & \\
U118 &  17:40:41.576  $-$53:40:15.88   &  0.05    & \nd             & 1         & MSTO   & c   & c    &  15.60  &  16.60  &  18.34 & only object in error circle \\
U119 &  17:40:41.261  $-$53:40:19.35   &  0.22    & \nd             & 0         & \nd    & \nd & \nd  &  \nd    &  \nd    &  \nd   & empty error circle \\
U120 &  17:40:46.517  $-$53:40:15.64   &  0.21    & \nd             & 1         & AB?    & ?   & ?    &  21.09  &  23.70  &  23.39 & moderately red, large Ha excess \\
U121 &  17:40:33.631  $-$53:39:34.96   &  0.68    & \nd             & 0         & \nd    & \nd & \nd  &  \nd    &  \nd    &  \nd   & empty error circle \\
U122 &  17:40:47.903  $-$53:39:24.83   &  1.55    & \nd             & 0         & \nd    & \nd & \nd  &  \nd    &  \nd    &  \nd   & empty error circle \\
U123 &  17:40:49.621  $-$53:38:45.93   &  2.04    & \nd             & 0         & \nd    & \nd & \nd  &  \nd    &  \nd    &  \nd   & empty error circle \\
\enddata			  
\tablenotetext{a}{From \citet{Bogdanov10}}
\tablenotetext{b}{Projected distance from cluster center}
\tablenotetext{c}{New IDs are indicated by an asterisk;
Previous CV IDs are from \citet{Grindlay06} and references therein;
Prefix BY (Dra) IDs are from \citet{Taylor01};
Prefix V (variable) IDs are from \citet{Kaluzny03} and \citet{Kaluzny06}}
\tablenotetext{d}{1 = object detected in error circle; 0 = no object detected in error circle}
\tablenotetext{e}{
CV = cataclysmic variable; CV? = less certain CV identification for reason
noted in table;
AB = active binary candidate; AB? = less certain AB identification
for reason noted in table; 
MSP = millisecond pulsar;
MS = main sequence; 
BS = blue straggler;
AGN = active galactic nucleus;
GLX = interacting galaxies
}
\tablenotetext{f}{Proper-motion membership using 2006 second epoch: c
  = consistent with cluster; f = consistent with field; ? = unknown
  membership status from proper-motion data; \nd\ = no proper-motion
  measurement}
\tablenotetext{g}{Proper-motion membership using 1999 second epoch.}
\end{deluxetable}
\clearpage
\end{landscape}

\end{document}